\documentclass[fleqn,usenatbib]{mnras}

\usepackage{color}
\usepackage{graphicx}
\usepackage{dcolumn}
\usepackage{bm}
\usepackage{amssymb}
\usepackage{latexsym}
\usepackage[T1]{fontenc}
\usepackage{aecompl} 
\usepackage{amsmath}%
\usepackage{enumitem}
\usepackage{epstopdf}

\newcommand{\be}{\begin{equation}}
\newcommand{\ee}{\end{equation}}
\newcommand{\bq}{\begin{eqnarray}}
\newcommand{\eq}{\end{eqnarray}}

\def\({\left(}
\def\){\right)}

\title[Cosmology from Gradient Weighted Correlation]{Cosmological constraints from the density gradient weighted correlation function}

\author[Xiao, Yang, Luo, et al. (2021)]{ 
Xiaoyuan Xiao$^{1,2}$,
Yizhao Yang$^{1,2}$\thanks{E-mail: yangyzh33@mail2.sysu.edu.cn}, 
Xiaolin Luo$^{1,2,\dagger}$\thanks{E-mail: luoxl23@mail2.sysu.edu.cn}, 
Jiacheng Ding$^{1,2,\dagger}$,
Zhiqi Huang$^{1,2}$, 
\newauthor
Xin Wang$^{1,2}$, 
Yi Zheng$^{1,2,\ddagger}$\thanks{E-mail: zhengyi27@mail.sysu.edu.cn},  
Cristiano G. Sabiu$^3$, 
Jaime Forero-Romero$^4$, 
\newauthor
Haitao Miao$^5$, 
Xiao-Dong Li$^{1,2,\star}$ \thanks{E-mail: lixiaod25@mail.sysu.edu.cn}
\\
$^1$ School of Physics and Astronomy, Sun Yat-Sen University, Guangzhou 510297, P.R.China \\
$^2$ CSST Science Center for the Guangdong-Hong kong-Macau Greater Bay Area, SYSU \\
$^3$ Natural Science Research Institute, University of Seoul, 163 Seoulsiripdaero, Dongdaemun-gu, Seoul, 02504, Republic of Korea \\
$^4$ Departamento de F{\'i}sica, Universidad de los Andes, Cra. 1 No. 18A-10 Edificio Ip, CP 111711, Bogot{\'a}, Colombia \\
$^5$ Key Laboratory for Computational Astrophysics, National Astronomical Observatories, Chinese Academy of Sciences,
20A Datun Road, Beijing 100012, People's Republic of China \\}
\date{Accepted XXX. Received YYY; in original form ZZZ}

\pubyear{2021}

\begin{document}
\label{firstpage}
\pagerange{\pageref{firstpage}--\pageref{lastpage}}
\maketitle

\begin{abstract}
The mark weighted correlation function (MCF) $W(s,\mu)$ 
is a computationally efficient statistical measure which can probe 
clustering information beyond that of the conventional 2-point statistics.
In this work, we extend the traditional mark weighted statistics 
by using powers of the density field gradient $|\nabla \rho/\rho|^\alpha$
as the weight, 
and use the angular dependence of the scale-averaged MCFs
to constrain cosmological parameters.
The analysis shows that the gradient based weighting scheme 
is statistically more powerful than the 
density based weighting scheme,
while combining the two schemes together is more powerful 
than separately using either of them.
Utilising the density weighted or the gradient weighted MCFs with $\alpha=0.5,\ 1$,
we can strengthen the constraint on $\Omega_m$ by factors of 2 or 4, respectively, compared with the standard 2-point correlation function,
while simultaneously using the MCFs of the two weighting schemes together can be
$1.25$ times more statistically powerful than using the gradient weighting scheme alone.
The mark weighted statistics may play an important role in cosmological analysis of future large-scale surveys.
Many issues, including the possibility of using other types of weights, 
the influence of the bias on this statistics, 
as well as the usage of MCFs in the tomographic Alcock-Paczynski method, 
are worth further investigations. 
\end{abstract}









\section{Introduction}\label{intro}

The cosmic acceleration \citep{riess1998observational,perlmutter1999measurements} 
indicates
either the presence of a ``dark energy'' component in the Universe
or the breakdown of general relativity on cosmological scales. 
In the last two decades, tremendous effort has been put on the theoretical explanation and observational probes of acceleratting universe,
yet they are still not well comprehended or accurately measured 
\citep{weinberg1989cosmological,miao2011dark,YOO_2012,weinberg2013observational}.


On large scales, the spatial distribution of galaxies 
forms a unique, and complex filamentary motif known as the 'cosmic web'
\citep{1986Bardeen,1986deLapparent,Huchra_2012,Tegmark_2004,Guzzo_2014},
which encodes a significant amount of information about the expansion and structure growth history of the Universe. 
Over the next few years, surveys of large-scale structure (LSS) like DESI\footnote{https://desi.lbl.gov/}, EUCLID\footnote{http://sci.esa.int/euclid/}, 
Rubin\footnote{https://www.lsst.org/},
Roman\footnote{https://wfirst.gsfc.nasa.gov/}, and CSST \citep{Gong_2019},
will explore an unprecedented large volume of the Universe with extraordinary precision. 
So the development of powerful statistics is of great significance to 
comprehensively and reliably derive the cosmological parameters
from LSS.

The most widely-adopted LSS statistics are the 2-point correlation function (2pCF) or power spectrum measurements
\citep{kaiser1987clustering,ballinger1996measuring,Eisenstein_1998,Blake_2003,Seo_2003}.
These methods have had excellent success when applied to
galaxy redshift surveys such as the 2-degree Field Galaxy Redshift Survey
(2dFGRS; \cite{2df:Colless:2003wz}),
the 6-degree Field Galaxy Survey (6dFGS; \cite{beutler20116df}),
the WiggleZ survey \cite{blake2011wigglez,blake2011wigglezb},
and the Sloan Digital Sky Survey (SDSS;
\cite{york2000sloan,Eisenstein:2005su,Percival:2007yw,
anderson2012clustering,sanchez2012clustering, sanchez2013clustering, 
anderson2014clustering, samushia2014clustering,ross2015clustering,
beutler2016clustering,sanchez2016clustering,
alam2017clustering,chuang2017clustering}).
However, the main defect of these methods is that they are only sensitive to 
the Gaussian statistical properties of the density field,
while both structure formation processes and the primordial conditions 
can introduce significant amounts  of non-Gaussianity into the LSS.



Ongoing studies seek to go beyond the 2-point statistics
explored methods like the 3-point statistics \citep{Sabiu_2016,Slepian_2017},
the 4-point statistics \citep{Sabiu_2019},
the density field or voids \citep{ryden1995measuring,lavaux2012precision,Sutter:2012tf,Bos:2012wq,Chan:2014qka,Cai:2014fma,Sutter:2014oca,
Qingqing2016,KR2018,Hamaus:2020cbu,Lavaux:2021ncp},
the Minkowski functionals \citep{Minkowski1903,Mecke:1994ax,Schmalzing:1997uc,Kerscher:1998gs,Park2010,Appleby:2020pem,Appleby:2021lfq,Appleby:2021xoz},
%
deep learning \citep{Ravanbakhsh17,Mathuriya18,He:2018ggn,Ntampaka:2019ole,Pan:2019vky,Li:2020vor,Mao:2020vdp,Villaescusa-Navarro:2020rxg,
Ni:2021mzk,WuZiyong:2021jsy},
and so on.
While all of them can explore the non-Gaussian clustering information encoded in the LSS,
in this analysis we investigate a statistical tool namely the {\em mark weighted correlation function} 
\citep[MCF; ][]{Beisbart2000,Beisbart2002,Gottl2002,Sheth:2004vb,Sheth:2005aj,Skibba2006,White_2009,
White_2016,Satpathy:2019nvo,Alam:2020jdv,PMS2020,Massara:2020pli}
which is simple and computationally efficient compared with the aforementioned statistics.

The MCF is simply the 2-point statistics but with each object weighted by a ``mark'' related with its local environment.
The weights are generally proportional to the local density to a positive or negative power.
In this way, the statistics give more emphasis on the dense or the undense regions.
This leads to better discrimination of the different patterns of galaxy clustering and redshift space distortions (RSDs) 
in the dense or under dense areas, and thus can lead to 
tighter cosmological constraints 
compared with the standard 2pCF \citep{Yang:2020ysv}.


\begin{figure*}
\centering
\includegraphics[width=20cm, trim=150 0 30 0,clip]{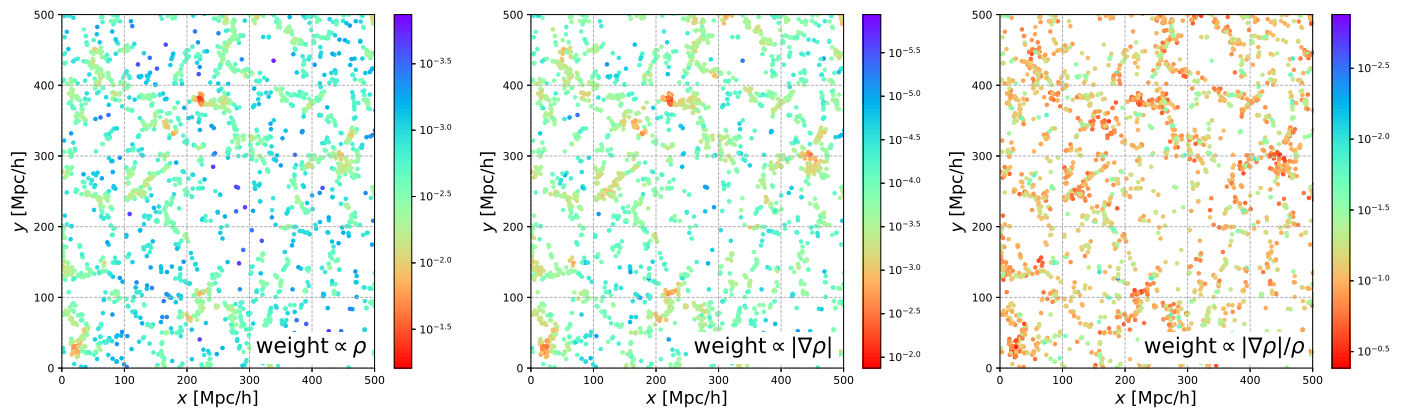}
\caption{ A slice of halos selected from the BigMDPL simulation.
Color of the points represents their weights in three weighting schemes.
{\it Left panel}: In the density weighting scheme, 
weight is solely determined by the local density.
Objects in the  regions of clusters (voids)
are assigned with very large (small) weights.
{\it Middle panel}: A scheme wherein the weights are set as $\nabla\rho$. 
In this case the weights are still crucially determined by the values of $\rho$. 
{\it Right panel:} The gradient weighting scheme we used in this analysis.
To eliminate the dependence on $\rho$ we set ${\rm weight} \propto |\nabla \rho|/\rho$.
This weighting is thus sensitive to the change of density in the local environment, 
while not being overly sensitive to its absolute value.
In this plot, values of the object weights in the voids and clusters 
have the same order of magnitude.
\label{fig:method}}
\end{figure*}

While using the local density is the simplest way to 
include the local environment information into the analysis, 
this approach is certainly far from capturing the full information 
of the environment.
In previous works, many statistical quantities related with the 
local environment have been proposed and studied 
\citep{Gott:1986uz,Mecke:1994ax,hahn2007properties,Gott:2008kk,forero2009dynamical,2011MNRAS.414..350S,hoffman2012kinematic,Park:2013dga,
Wang:2013gtn,forero2014cosmic,Pranav:2018pnu,Fang2019,Hong:2019qix,Garcia-Alvarado:2020wvx,Shim:2020qav,Shim:2020wyj,Suarez-Perez:2021dqs}.
So it may be worth considering other statistical quantities 
in the construction of the MCFs,
so as to extract more information from the LSS. 

In this paper, we extend the scope of the marked statistics 
by considering more possible weighting in the correlation functions. 
Tentatively, in this analysis we try using the {\it density gradient} as the mark,
based on three considerations.
1) The first order derivative is the simplest extension to the value itself.
It is natural to consider the spatial variation of the field 
after its value being considered.
2) The density can be used to classify the universe
into topological regions of cluster, filament, sheet and void.
In contrast, the gradient traces regions where the density changes rapidly,
and thus marks the boundaries among the different topological regions.
From this viewpoint, the density and the density gradient are complementary descriptions of the LSS.
So they could also be complementary in a certain statistics.
3) Phenomenologically, regions having large density gradient are expected to 
have large gravitational acceleration.
Since velocity is not directly observable in surveys, 
density gradient could be the simplest and most convenient quantity 
if one were to study the dynamical property of the LSS.

This paper is organized as follows.
Section \ref{data} describes the simulation data used in this analysis. 
In Section \ref{methodology}, we explained the concept and the mathematical definition of the marked statistics we designed. 
In Section \ref{results}, we applied the marked statistics to the simulation data, and presents the obtained cosmological results.
We discuss and conclude in Section \ref{conclusion}.

\begin{figure}
	\centering
	\includegraphics[width=8cm]{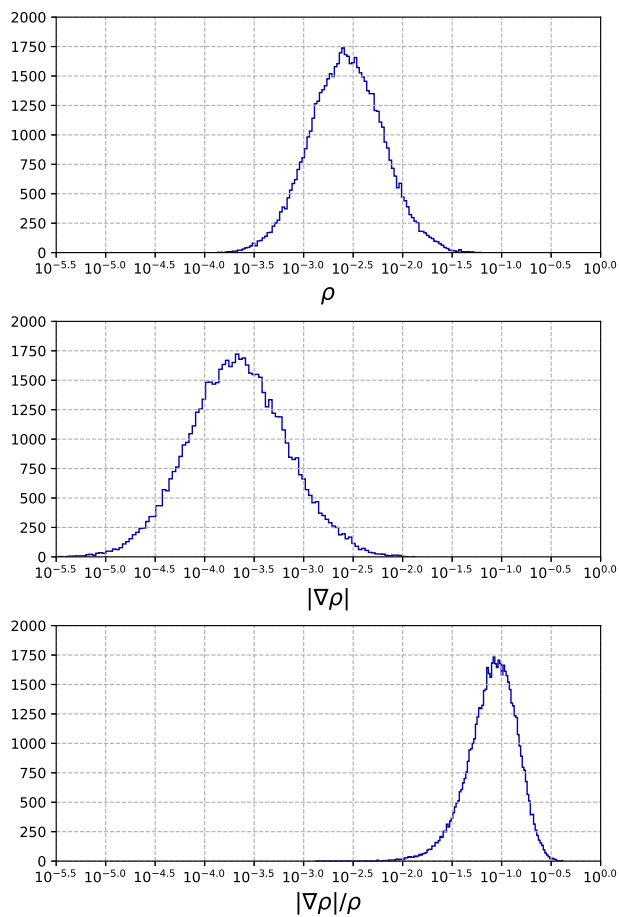}
\caption{
Histograms of the object weights defined as 
$\rho$, $|\nabla \rho|$ and $|\nabla \rho|$/$\rho$, respectively.
The distributions of the former two schemes are quite similar to each other.
The last weighting scheme exhibits a clearly different distribution,
and the dispersion of the values is significantly smaller compared with the other two schemes. 
\label{fig:loghist}}
\end{figure}


\section{DATA}\label{data}

To conduct a cosmological analysis using the MCFs of the two weighting schemes,
we use the BigMulti-Dark\footnote{webpage: https://www.cosmosim.org } (BigMD) N-body simulation, 
and also a set of samples produced using the COLA (COmoving Lagrangian Acceleration) algorithm \citep{Tassev13}.
In all simulations, we identify gravitationally bound structures using ROCKSTAR \citep{ROCKSTAR},
which is a halo finder based on adaptive hierarchical refinement of friends-of-friends groups in six phase-space dimensions and one
time dimension to guarantee robust tracking of substructure.
We fixed a halo number density $\bar n= $ 0.001 $(h^{-1}\rm Mpc)^{-3}$ in all simulations to ensure 
the comparability.
Both halos and subhalos are used in our analysis.

\subsection{BigMD simulation}

The Big MultiDark (BigMD) Planck simulation belongs to the series of MultiDark simulations.
It is generated using $3\,840^3$ particles in a volume of $(2.5h^{-1}\rm Gpc)^3$,
based on a flat $\Lambda$CDM cosmology with Planck parameters
$H_0 = 67.77\ {\rm km}\ s^{-1} {\rm Mpc}^{-1}$, $\Omega_m = 0.307115$, $\Omega_b = 0.048206$, $n_s = 0.9611$, 
and  $\sigma_8 = 0.8288$ \citep{BD}.
The initial condition was set
by using the Zeldovich approximation at redshift $z_{\rm init} = 100$.
Both the volume and the number of particles are very huge,
so this simulation is among the best choices for testing our methodology.

\subsection{COLA MultiCosmo Simulations}

In order to study the sensitivity of the statistics to cosmological parameters,
we also generate five COLA simulations using different cosmological parameters of 
$\Omega_m = 0.2671,\ 0.2871,\ 0.3071,\ 0.3271$ and $0.3471$, respectively.
Values of the other parameters, i.e. $\Omega_b$, $\sigma_8$, $n_s$ and $H_0$, 
are fixed to the values adopted in the BigMD simulation. 
To minimize the effect of cosmic variance, 
we maintain the same random seed when generating their initial conditions.
These simulations are produced using $1024^3$ particles in a box of 
$(680 h^{-1}\rm Mpc)^3$. 
For convenience, in what follows we refer to them as {\it COLA MultiCosmo Simulations}.

COLA combines 2LPT (the second order Lagrangian perturbation theory), for time integration
for large scale dynamical evolution, with a full-blown N-body Particle-Mesh (PM) algorithm to calculate the small scale dynamics \citep{Tassev13}.
In this work, COLA was chosen because it is hundreds of times faster than N-body simulation and can still maintain a good accuracy on non-linear scales.

\subsection{Simulations for Covariance Estimation}

To compute the covariance matrices we generate 
180 simulations using COLA.
All simulations are produced using $600^3$ particles in a $(512 h^{-1}\rm Mpc)^3$ box,
and their cosmological parameters are the same as those adopted in the BigMD simulation.
The only difference of these simulations is that we use different random seeds 
in creating their initial conditions.

\section{Methodology}\label{methodology}

As a simple generalization of the standard 2-point correlation function,
MCF assign each object an environment dependent weight,
which is commonly chosen as a form proportional to the local density \cite{White:2016yhs,Yang:2020ysv} 
\begin{equation}\label{eq:weightrho}
{\rm weight}=\rho_{n_{\rm NB}}^\alpha,
\end{equation}
Here we estimate the local density $\rho_{n_{\rm NB}}$ using its $n_{\rm NB}$ nearest neighbors, i.e.
\begin{equation}
 \rho_{n_{\rm NB}}({\bf r}) = \sum_{i=1}^{n_{\rm NB}}  W_k({\bf r-r_i},h_W),
\end{equation}
where $\rho_{n_{\rm NB}}(\bf r)$ is the environmental number density around a specific galaxy located at $r$,
and $W_k$ is the smoothing kernel, for which we choose the 3rd order B-spline functions
having non-zero value within a sphere of radius $2h_W$ $h^{-1}$ Mpc
\citep{Gingold1977, Lucy1977}.
In order to ensure that the kernel always contains $n_{\rm NB}$ 
nearest neighbor halos within a radius of $2h_W$,
We use an adjustable radius for the smoothing kernel.


As the simplest generalization of Equation \ref{eq:weightrho},
in this analysis we use the gradient of the local density as the weight, and define 
\begin{equation}\label{eq:gradient_weight}
{\rm weight}=(|\nabla \rho_{n_{\rm NB}}|/\rho_{n_{\rm NB}})^{\alpha},
\end{equation}
where the gradient is evaluated via the derivatives of the kernel,
\begin{equation}
 \nabla \rho_{n_{\rm NB}}({\bf r}) = \sum_{i=1}^{n_{\rm NB}} \nabla W_k({\bf r-r_i},h_W).
\end{equation}
Equation \ref{eq:gradient_weight} describes the variation of the local density around a certain position.
In order to focus on the variation of $\rho$, rather than being sensitive to its value,
we divide the value of the gradient by the value of $\rho$.
Throughout this analysis, we choose $n_{\rm NB}=30$,
which leads to a smoothing radius $h_W=8.3\pm3.7 h^{-1} \rm Mpc$.

For simplicity, we will refer the above two weighting schemes
as {\it density weighting scheme} and {\it gradient weighting scheme},
respectively.
We will apply both of them to the simulation data 
and compare the cosmological results obtained.

As an illustration of the weighting schemes,
Figure \ref{fig:method} shows a slice of objects selected from the BigMD simulation, 
with weights of the objects represented by their colors.
In the density weighting scheme, 
objects in the most dense regions (clusters) are assigned with very large weights,
while objects in void regions are assigned with negligible weights.
In case that we use the gradient, and assign each object a ${\rm weight}$ defined by $\nabla\rho$, 
the results are still quite similar to the results of the density weighting scheme,
The reason is that the value of $|\nabla \rho|$ is mostly determined by the value of $\rho$ itself.
Finally, the right panel represents the weighting scheme used in this analysis, 
where we use ${\rm weight} \propto |\nabla \rho|/\rho$,
which is not too dependent on $\rho$ while still sensitive to the variation of the environment.

The distribution of the weights are illustrated in Figure \ref{fig:loghist}.
The distribution of the $\rho$ and $|\nabla\rho|$ weighting schemes are similar.
For both of them, we detect 4-5 orders of magnitude variation among the majority of objects.
In contrast, the $|\nabla \rho / \rho|$ weighting scheme displays a clearly different distribution,
and the dispersion of the objects' weights is only $\sim$1-2 magnitude.
We consider the low dispersion to be a particularly advantageous property of this weighting scheme.


Except the difference in the weighting,
the other process of the statistics are exactly the same as those in general MCFs.
In the measurement of the 2pCF,
we adopt the most commonly used Landy-Szalay estimator 
\begin{equation}\label{eq:Wsmu}
W(s,\mu) = \frac{WW(s,\mu)-2WR(s,\mu)+RR(s,\mu)}{RR(s,\mu)}.
\end{equation}
Here $WW(s,\mu)$ denotes the weighted number of galaxy-galaxy pairs,
$WR(s,\mu)$ presents the number of galaxy-random pairs,
and $RR(s,\mu)$ corresponds to the count of random-random pairs. 
The pairs separation of objects are defined by 
$s\pm \Delta s$ and $\mu\pm \Delta \mu$,
where the letter $s$ is the distance between the pair of objects in the redshift space and
$\mu=\cos(\theta)$, where $\theta$ denotes the angle 
between the line joining the pair and the line of sight direction
\footnote{We use $s$ instead of $r$ here, because the statistics is usually computed using the distance in redshift space, 
	and due to the RSDs they are correlated with each other via $s = r+v/(aH)$.}.
We always use 10 times more particles in a random samples
than in the data samples.

\begin{figure*}
	\centering
	\includegraphics[width=16cm]{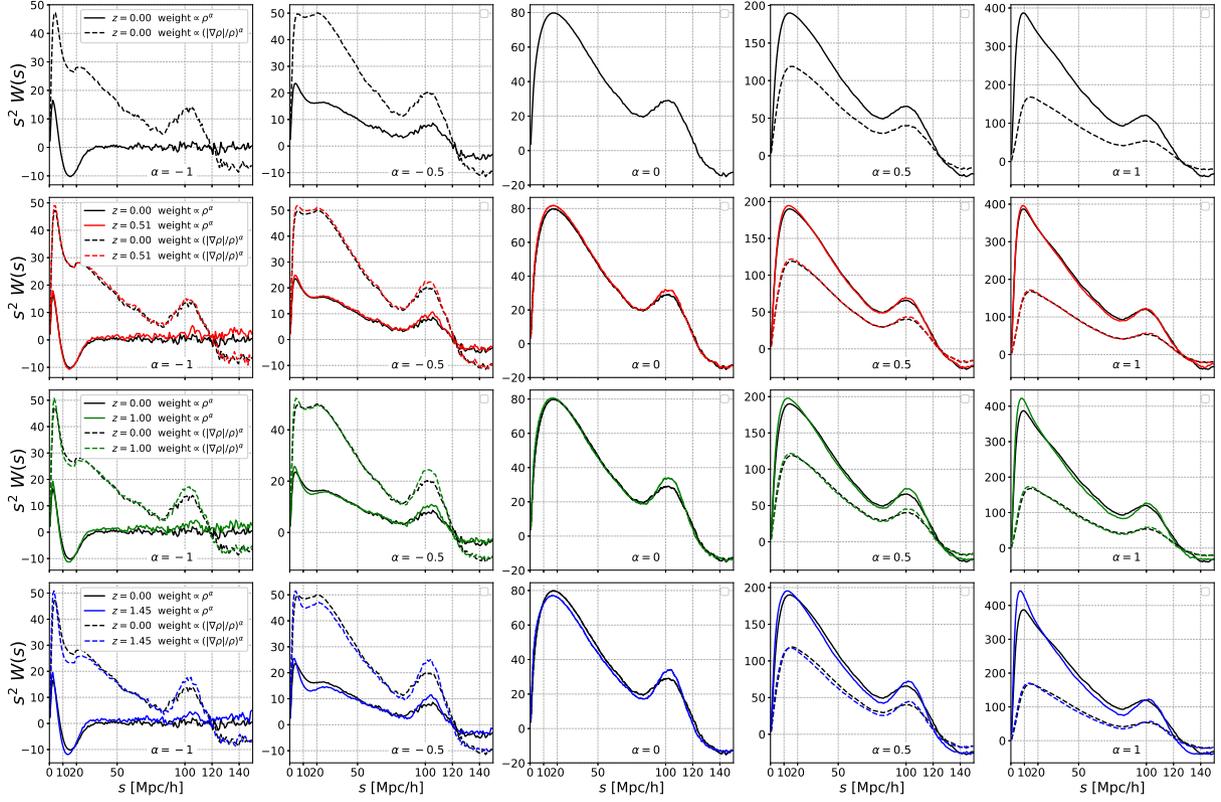}
\caption{
The monopole measurements $s^2 W(s)$ in the density and gradient weighting schemes. 
Values of $\alpha$ are set as $-1,\ -0.5,\ 0,\ 0.5$ and $1$, respectively.
The uppermost panels show the results at $z=0$,
while panels in the other rows compare the $z=0.51$, 1.0, 1.45 results to the redshift zero results. 
The two schemes show largest difference in the case of $\alpha=-1$,
In the density weighting scheme, the amplitude of the correlation function
increases faster as we increase $\alpha$.
\label{fig:Ws}}
\end{figure*}

\begin{figure*}
	\centering
	\includegraphics[width=16cm]{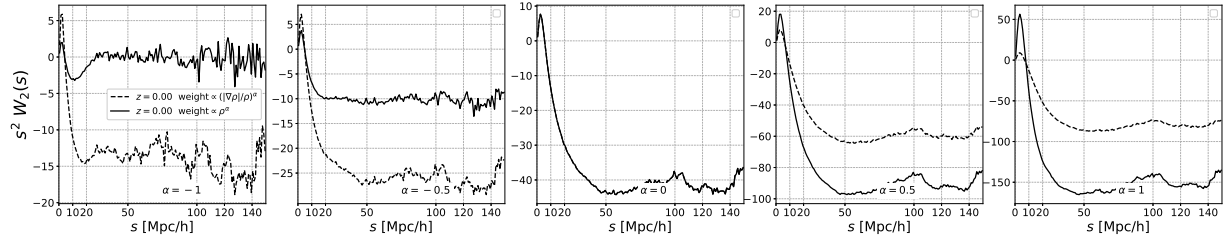}
\caption{
The same to Figure \ref{fig:Ws}, but for the $s^2 W_2(s)$ measurements at $z=0$.
\label{fig:W2s}}
\end{figure*}

In the counting of pairs, the weighting scheme is applied to both 
the data and the random samples 
(except that we use a 10x larger $n_{\rm NB}$ in the random).
This treatment is different from what adopted in \citep{Yang:2020ysv}
(where the authors fixed the weights of the randoms as 1).
Here we do this to take into consideration the possible 
correlations between nearby objects induced by the smoothing,
although we tested and found that there are no significant changes to the measured 2pCFs.

\section{Results}\label{results}

In what follows we present our results.
Sec. \ref{subsec:Ws} presents the monopole $W(s)$ and quadrupole $W_2(s)$ of the two weighting schemes.
Sec. \ref{subsec:Wmu} presents the anisotropic $W_{\Delta s}(\mu)$ measurements.
Their covariance matrices are presented in Sec. \ref{subsec:Cov}.
We derive the cosmological constraints in Sec. \ref{subsec:Cosmology}.


\subsection{The angular-averaged measurements}\label{subsec:Ws}

\begin{figure*}
	\centering
        \includegraphics[width=16cm]{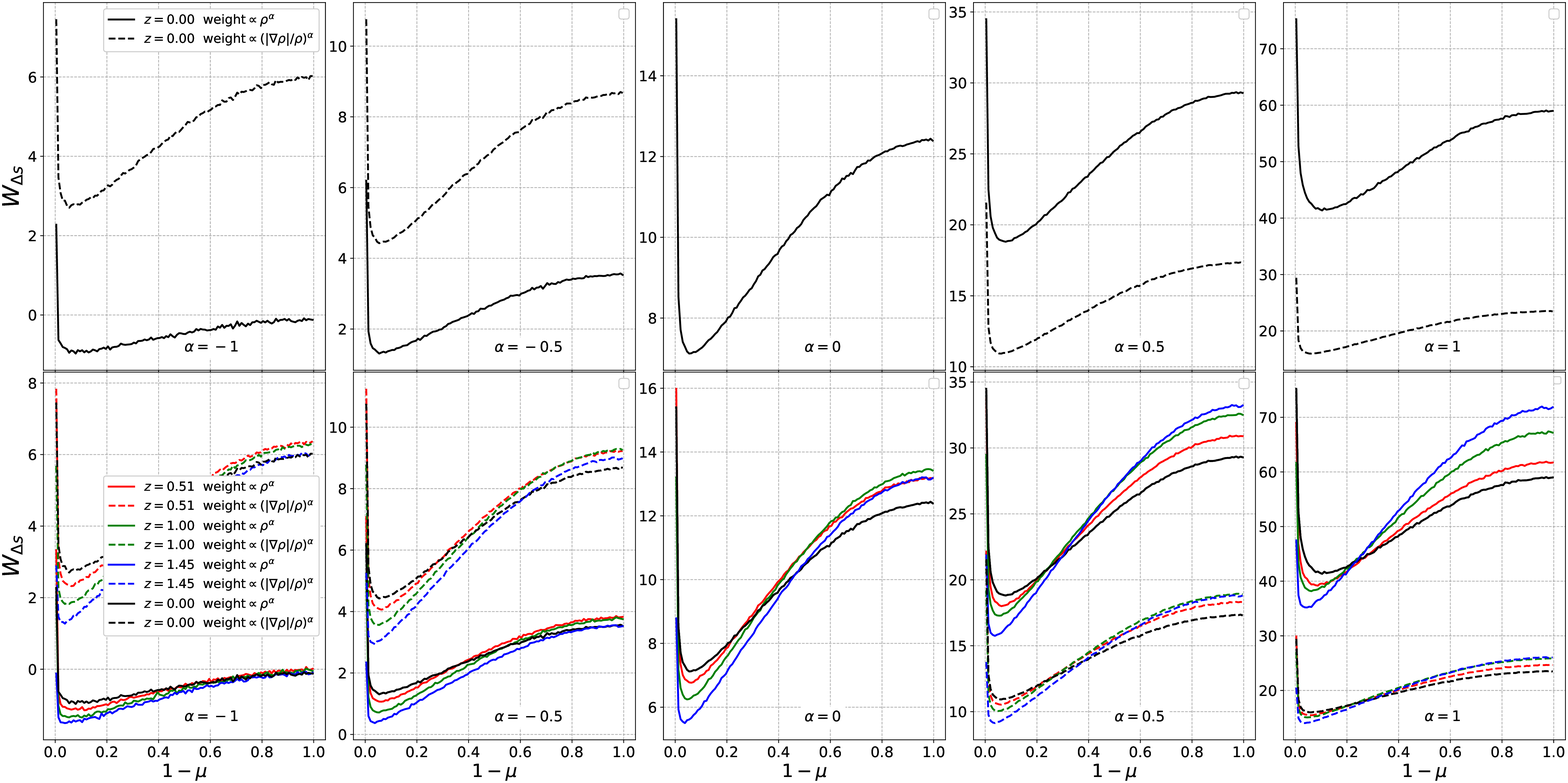}
\caption{
The scale-averaged statistics $W_{\Delta s}(\mu)$ measured in the two weighting schemes.
The slope decreases as we go to higher redshifts where the RSD effect becomes weaker. \label{fig:Wmu}
}
\end{figure*}

In this subsection we present the angular-averaged MCFs (i.e. the multipoles), which are defined as 
\begin{equation}
 W_l(s)\equiv \frac{2l+1}{2}\int_{-1}^{1} L_l(\mu) W(\mu, s)d\mu.
\end{equation}
In what follows, we present the monopole and quadrupole measurements.
For simplicity, we will use $W(s)$ rather than $W_0(s)$ to denote the monopoles.

Figure \ref{fig:Ws} shows the  monopole measurements $s^2 W(s)$ obtained using the 
density and gradient weighting schemes,
with $\alpha$ taken as -1, -0.5, 0, 0.5 and 1, respectively.
Each measurement uses a halo/subhalo sample identified from one BigMDPL snapshot, 
with a fixed number density $10^{-3} (h^{-1} \rm Mpc)^{-3}$ produced by applying a minimal cut of halo/subhalo mass.
As described in the previous section, values of $\rho$ and $\nabla \rho$ are estimated using 30 nearest neighbors,
corresponding to a smoothing radius of $h_W=8.3\pm3.7 h^{-1} \rm Mpc$.


The first row shows the results measured at $z=0$.
From left to right, we see a clear trend of increasing
clustering strength as we increase the value of $\alpha$,
meaning that in regions with larger $\rho$ or $|\nabla \rho| / \rho$ 
objects are more clustered.
But the trend is less significant in the gradient weighting scheme.
The shapes and amplitudes of the $s^2W(s)$s are affected by the choice of the weighting schemes.

The largest difference between the two schemes is 
detected at the $\alpha=-1$ case.
Effectively, weighting the objects by $\rho^{-1}$ is equivalent to transforming the distribution to a spatially uniform distribution.
So in the density weighting scheme we find $W(s)\approx 0$ at $s\gtrsim 20 {h^{-1}} \rm Mpc$
\footnote{Considering that the radii of the smoothing sphere is $2h_W=16.6\pm7.4 {h^{-1}}\rm Mpc$,
clustering signal above this scale is is erased.
Below this scale, the smoothing kernel can
induce some ``weird'' clustering features  (the sharp peaks and valleys 
in $\alpha=-0.5$ and $\alpha=-1$ cases),
which is discussed in \cite{Yang:2020ysv}  and is
found to be related with the scale of the smoothing.}.
Of course, this phenomenon is not detected if weighting is $|\nabla \rho/\rho|^{-1}$.

Figure \ref{fig:Ws} also compares the $W(s)$ measurements at different redshifts.
In all cases, we observe a similar amplitude at different redshifts,
and a more prominent BAO peak at higher redshifts.

Finally, Figure \ref{fig:W2s} shows the quadrupole measurements at $z=0$.
Similar to the monopole results, 
here we see the two weighting schemes lead to measurements having significantly different shapes and amplitudes,
and the density weighting scheme measurement varies more dramatically when we change the value of $\alpha$.



\subsection{The scale averaged measurements $W_{\Delta s}(\mu)$}\label{subsec:Wmu}

The dynamics of objects crucially depends on their environment.
This results in an environmental dependent RSD,
and motivates us to study the anisotropic clustering using the mark weighted correlation functions.
To do this, we measured the scale-averaged correlation function $W(s,\mu)$,
which quantifies the anisotropy via 
\begin{equation}
 W_{\Delta s}(\mu) \equiv \int_{s_{\rm min}}^{s_{\rm max}} W(s, \mu) ds.
\end{equation}
and its normalized version
\begin{equation}
 \hat W_{\Delta s}(\mu)\equiv \frac{W_{\Delta s}(\mu)}{\int_{0}^{\mu_{\rm max}}W_{\Delta s}(\mu)\ d\mu}.
\end{equation}
These two quantities were used to quantify the RSDs and the Alcock-Paczynski (AP) distortions
in the tomographic AP method \citep{LI15,Park:2019mvn,Ma_2020}, 
and have been proved useful in constraining the dark matter ratio $\Omega_m$ \citep{LI16}, 
the dark-energy equation of state parameter $w$ \citep{LI16} and its time evolution \citep{LI18, Zhang2019},
the Hubble constant \citep{Zhang:2018jfu},
the curvature and the properties of neutrinos \citep{LI19}.

Figure \ref{fig:Wmu} present the measured $W_{\Delta s}(\mu)$s
when $s_{\rm min},\ s_{\rm max}$ are set as  6, 40 $h^{-1}\rm Mpc$, respectively.
In all cases, we detect a sharp peak at $0\lesssim1-\mu\lesssim0.1$,
and a slope in the range of $0.1\lesssim1-\mu$ 
The former is produced by the finger-of-god (FOG) effect \citep{Jackson},
while the latter is the result of the Kaiser effect \citep{kaiser1987clustering}.

\cite{Yang:2020ysv} showed that $\hat{W}_{\Delta s}(\mu)$ 
has a statistically stronger constraint on cosmological parameters than $W(s)$.
Also, in the construction of $\hat W_{\Delta s}$ we integrate the scale-dependence and normalize the amplitude,
so compared with $W(s)$ it is less likely to be affected by scale-dependent bias.
So in the following analysis we only use $\hat{W}_{\Delta s}(\mu)$. 

\subsection{Covariance measurements}\label{subsec:Cov}

\begin{figure}
	\centering
	\includegraphics[width=8cm]{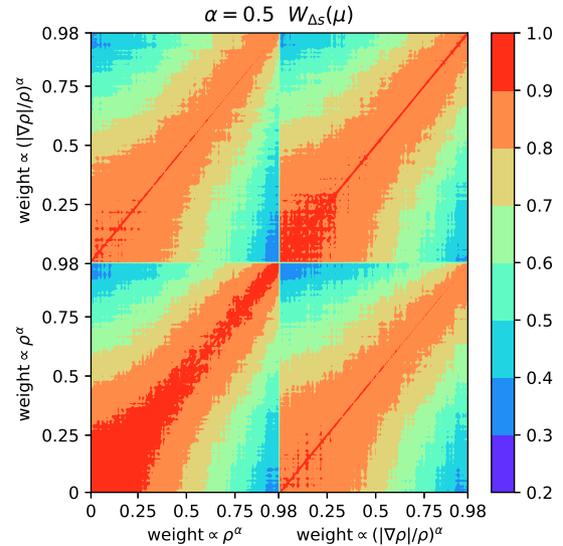}
	\caption{
	Correlation coefficient of $W_{\Delta s}(\mu)$.
	In both of the two weighting schemes there exist positive correlation among nearby $\mu$-bins.
	Compared with the density weighting scheme, in the gradient weighting scheme the correlation among nearby $\mu$-bins is weaker.
	Strong cross-correlation between the two schemes is detected.
	\label{fig:coef}}
\end{figure}

\begin{figure*}
	\centering
	\includegraphics[width=12cm]{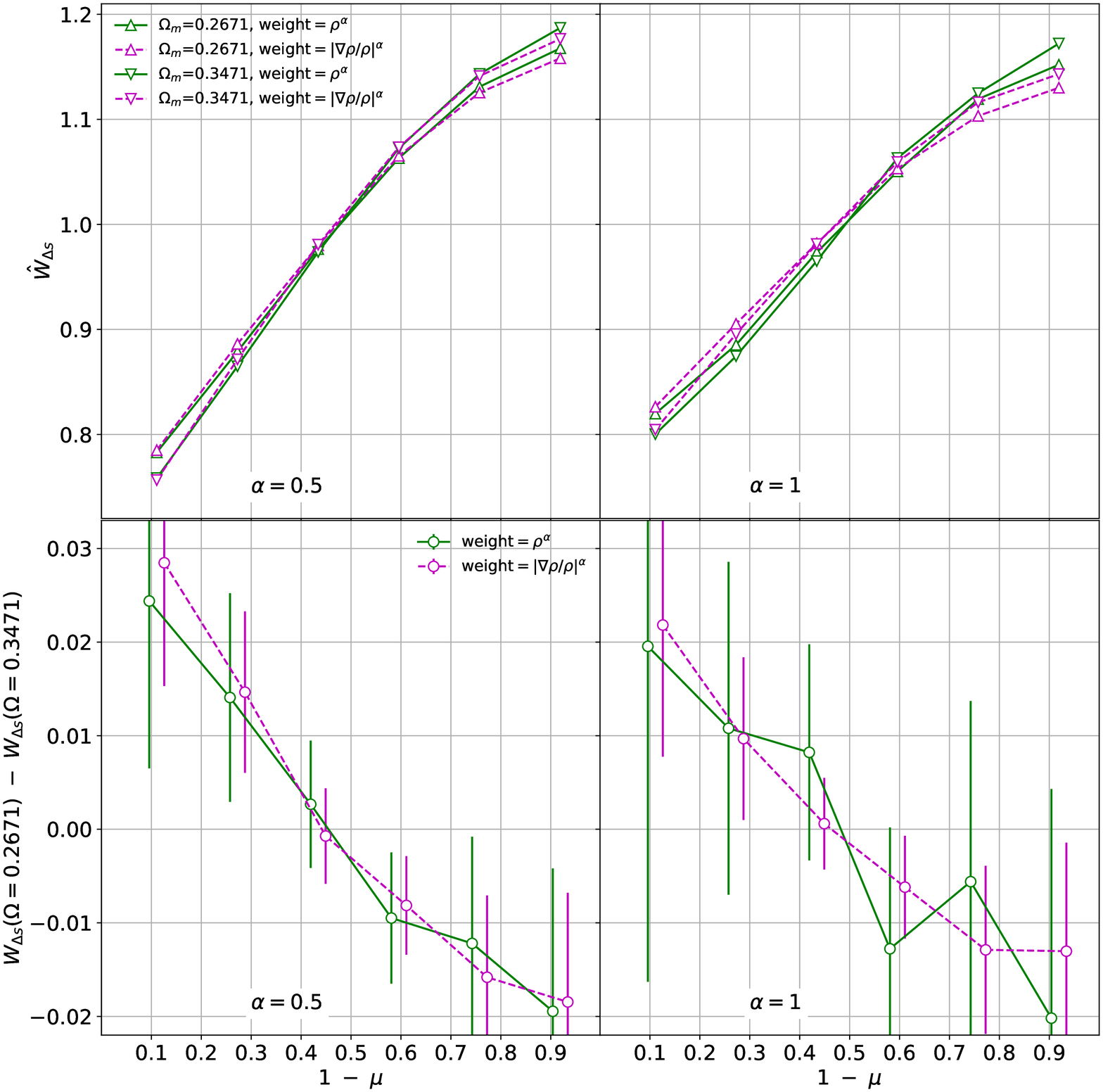}
	\caption{Using the MCFs to distinguish two different cosmologies with different values of $\Omega_m$.
	Upper panels show the measured $W_{\Delta s}(\mu)$,
	 while lower panels show the difference between the two cosmologies.
	In the left and right panels, we show the results obtained using $\alpha=0.5$ and 1, respectively.
	The gradient weighting scheme has smaller statistical errors compared with
	  the density weighting scheme.
	\label{fig:fig_ximu_diffcosmo}}
\end{figure*}

Figure \ref{fig:coef} shows the correlation coefficients of the 
$W_{\Delta s}(\mu)$ in 120 $\mu$-bins.
Without loss of generality we only present the results using $\alpha=0.5$.

In both of the weighting schemes 
there exist positive correlation among the nearby $\mu$-bins,
although in the gradient weighting scheme this correlation is slightly weaker.
Although measures have been taken to eliminate the similarity between the two weighting schemes,
their $W_{\Delta s}(\mu)$ measurements are still far away from being independent.
We find the cross-correlation coefficient $\gtrsim0.8$ for the near diagonal terms.
So we must to take the cross-correlation into consideration 
when simultaneously using the two weight schemes.


\subsection{Cosmological constraints}\label{subsec:Cosmology}

In what follows, we study whether the new weighting scheme
is helpful in improving the cosmological constraints.
To do this, we try to distinguish 
the COLA MultiCosmo Simulations using MCFs with different weighting schemes, 
and evaluate their statistical power.
The COLA MultiCosmo Simulations are created using values of 
$\Omega_m=0.2671, 0.2871, 0.3071, 0.3271$ and $0.3471$, respectively. 
A varying mass-cut is applied to create a series of $\bar n=10^{-3}\ (h^{-1}\rm Mpc)^{-3}$ halo/subhalo samples.
For simplicity, the $\Omega_m=0.3071$ simulation is chosen as the baseline, and all simulations are compared with it.
The difference is characterized by
\begin{equation}
 \chi^2 = \Delta  {\bf p}^{\rm T} \cdot {\bf Cov}^{-1} \cdot {\bf \Delta p},
\end{equation}
where $\Delta \bf p$ is defined as 
\begin{equation}
 \Delta {\bf p} \equiv {\bf p}(\Omega_m=0.3071) - {\bf p}(\Omega_m)
\end{equation}
and $\bf p$ refer to one or several MCFs,
whose weighting scheme and weighting power depends on our choice.
Throughout the analysis, we use $n_{\mu}=6$ bins for a single $\hat W_{\Delta s}(\mu)$ measurement.
This means that if combining together 2 or 4 MCFs we will have 12 or 24 statistical variables.
These numbers are safely small compared with the number of mocks (180) used to compute the covariance matrix \citep{Hartlap:2006kj}.

\begin{figure*}
	\centering
	\includegraphics[width=18cm]{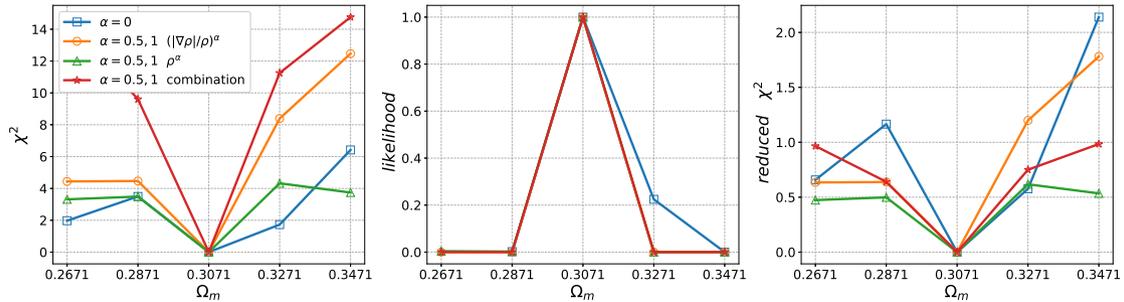}
	\caption{Using different MCFs to distinguish simulations with different cosmologies.
	The baseline cosmology has $\Omega_m=0.3071$, and the other cosmologies are compared to it.
	The $\chi^2$ and likelihood distributions of the cosmologies are plotted in the left and middle panels, respectively.
	Clearly, the gradient weighting scheme leads to tighter constraint than the density weighting scheme,
	while combining the two schemes together leads to
	a constraint tighter than solely 
	using any of them. 
	From left to right, we show the $\chi^2$, likelihood and reduced $\chi^2$, respectively. \label{fig:chisq}}
\end{figure*}

Figure \ref{fig:fig_ximu_diffcosmo} presents the $W_{\Delta s}(\mu)$ measured in two different cosmologies.
In all plots, we impose a cut $\mu>0.97$ to eliminate the clustering regimes severely affected by the FOG 
\footnote{This eliminates the sharp peak in $W_{\Delta s}(\mu)$ near the $1-\mu=0$ side; see \citep{LI15,LI16} for details.}.
In both the density and gradient weighting schemes, we are able to distinguish the two cosmologies using the MCFs.
In the lower panels (where we show the different between the MCFs in the two cosmologies), 
$\Delta {\bf p}$ of two weighting schemes does not coincide.
In what follows, we will consider simultaneously using them in the analysis and check 
whether combining them together is helpful 
in tightening the cosmological constraints. 


The results are illustrated in Figure \ref{fig:chisq}.
The left and middle panels show the distributions of the $\chi^2$s and the likelihoods $\mathcal{L}\propto e^{-\frac{1}{2}\chi^2}$,
respectively.
The plots include the results obtained using the standard 2pcf ($\alpha=0$), 
two density weighting scheme MCFs using $\alpha=0.5,\ 1$,
two gradient weighting scheme MCFs using $\alpha=0.5,\ 1$,
and the combination of the above four MCFs, respectively 
\footnote{We do not use the $\alpha=-0.5$ and -1 results which are found to be statistically weaker.}.
Since we are using the covariance estimated using the 180 COLA simulations,
all results represent the statistical power of 
a $\bar n=10^{-3} (h^{-1} {\rm Mpc})^{-3}$ halo/subhalo sample in a $(512 h^{-1} {\rm Mpc})^3$ volume.

Amazingly, we find that the gradient weighting scheme is statistically more powerful than the density weighting scheme.
Taking the $\Omega_m=0.3271$ cosmology as an example.
The standard 2pCF yields $\chi^2\approx1.3$, 
which means that we can hardly detect the difference between this cosmology and the baseline cosmology.
If using the two MCFs with ${\rm weight}\propto\rho^{0.5},\ \rho$,
the statistical significance is increased and we get $\chi^2\approx 2.9$.
In contrast, in the gradient weighting scheme, 
we get $\chi^2\approx 6$, which is almost twice as large as the former results.
Finally, combining the two weighting schemes yields $\chi^2\approx 8$,
which is better than using either of the schemes alone,
and is almost 5 times larger than the $\chi^2$ obtained using the standard 2pCF.

To take into consideration the degree of freedom (dof), the rightmost column shows the reduced $\chi^2$, 
which is defined as the $\chi^2$ divided by the dof
\footnote{As a not too bad approximation, here we define the dof as the number of points in the curve minus one (the number of parameters),
although the dof should be smaller due to the non-gaussian correlation among the points in the curve.}.
When looking at the values of the reduced $\chi^2$,
the results of the density weighting, gradient weighting, or their combination 
are not too different from each other.
This means that, the enlarged $\chi^2$ in the combined scheme is mainly due 
to the enlarged size of the data vector.
Also, compared with the results of the baseline cosmology,
the reduced $\chi^2$ in the wrong cosmologies are enlarged by 0.5-1.5,
meaning that they are strongly disfavored by the fitting
\footnote{Since the $\Omega_m=0.3071$ is chosen as the baseline 
here it perfectly ``fits'' the noise and leads to a zero reduced $\chi^2$.
This will not happen when we were analyzing the real observational data.}.

In summary, compared with the standard correlation function,
the density weighting or gradient weighting MCFs (with $\alpha=0.5$ and $1$)
are roughly 2 or 4 times more powerful in constraining $\Omega_m$,
while simultaneously using the two weighting scheme is $\approx1.25$ times more powerful 
than using only the gradient weighting scheme. 
The gradient weighting scheme is not only more powerful than the density weighting scheme,
but also is statistically complementary to it.

Finally, we mention three caveats to the above analysis.
1) We are limited by computational resources, in comparing cosmologies
we use five COLA simulations created by $1024^3$ particles in a $(680 h^{-1}{\rm Mpc})^3$ box.
Considering that we were using clustering at scales of $6-40 {h^{-1}}\rm Mpc$, 
the volume of the simulations should be large enough for the statistics.
However larger, and more accurate simulations are necessary if we are to achieve precise modelling of the cosmological dependence of $W_{\Delta s}(\mu)$.
2) $\hat {W}_{\Delta s}(\mu)$ may also not be the optimal choice of statistical quantity. 
This issue is worth further investigation.
3) Our analysis only considers $\alpha=0.5$ and 1,
so the conclusion is only valid if one were using these values of power.


\section{Conclusions and Discussions}\label{conclusion}


The MCFs (or the mark weighted power spectrum) can explore the 
clustering information not contained in the standard 2-point statistics,
while also being much more computationally efficient than three point or higher order statistics.
It is of ultra importance for us to studying them,
since they may play an important role in the analysis of future surveys.

In this work we consider using the power of the density gradient $(|\nabla \rho|/\rho)^\alpha$ 
as the weight in the marked two point statistics.
We use the anisotropic $\hat W_{\Delta s}(\mu)$ to distinguish simulations with different values of $\Omega_m$.
The $\chi^2$ and reduced $\chi^2$ shows that both weighting schemes 
are able to distinguish the different simulations.
The $\chi^2$ values indicate that,
the gradient weighting scheme is found to be statistically more powerful than the density weighted scheme,
while combining the two schemes together can further improve the statistical power.

The work can be considered as an extension of \cite{Yang:2020ysv} and \cite{Massara:2020pli},
where the authors found that the density weighting scheme can significantly 
increase the amount of statistical information of the standard two point analysis.
We show that the gradient weighted scheme can be complementary and even more powerful.


Phenomenologically it is not difficult to understand why the MCF is more powerful than the standard 2pCF. 
The clustering in different environments is different.
While the information from these regions are mixed together in the standard 2pCF,
they are separated if a weighting is applied, 
enabling us to extract more information.
However, it is much more difficult to explain 1) why the gradient weighting scheme is statistically more powerful than 
the density weighting scheme and 2) why the two weighting schemes are statistically complementary to each other.
Answering these two questions require accurate theoretical modeling of the statistical quantities
(e.g. the work of \cite{PMS2020}), and is beyond the scope of this work.

Although this proof-of-concept work shows that the density weighted correlation functions 
are helpful in the LSS analysis,
there is still a long way to go before we apply them to real observational data.
Here we list two caveats.
1) Weighted by the density or the density gradient, the mark weighted statistics could 
be more affected by the bias or the velocity bias of the tracers.
2) If one were to combine many MCFs together, 
a large amount of mocks are required to compute the covariance matrix.
In this case, one may have to use methods such like the principle component analysis 
to reduce the degrees of freedom. 

It is difficult to accurately calculate the MCFs from fundamental theory, hindering the comparison between theory and observations.
One possibility would be to run a large number of simulations and build up emulators for the MCFs.
An alternative option is to use the MCFs in the tomographic AP method \citep{LI14,LI15,LI16}.
The $W_{\Delta s}(\mu)$ with $\alpha=0$ is exactly the statistical quantity used in that method,
and so it is very likely to improve the power of that method via using more marked statistics.
We will explore this issue in future works.

There are still many issues related with the MCFs which are important but not addressed in this simple work.
As a preliminary study, here we only focus on the constraints of $\Omega_m$,
though it is worth exploring other physics such as modified gravity theories, neutrino mass and species, 
and the dark energy equation of state.
Also, while this work shows that one can improve the statistical power of the MCFs by using the density gradient,
it is known that this gradient encodes geometric information of the local cosmic web environment \citep{NovikovColombiDore}.
This could motivate further study of other cosmic web scalars as markers.
Some possibilities include the hessian of the potential field \cite{forero2009dynamical}, the hessian of the density field \citep{BondStraussCen}, the tidal anisotropy \citep{Bustamante} or even graph metrics \citep{Suarez-Perez:2021dqs},










\section*{Acknowledgements}


The CosmoSim database used in this paper is a service by the Leibniz-Institute for Astrophysics Potsdam (AIP).
The MultiDark database was developed in cooperation with 
the Spanish MultiDark Consolider Project CSD2009-00064.

This work is supported by National SKA Program of China No. 2020SKA0110401,
and the science research grants from 
the China Manned Space Project with No. CMS-CSST-2021-A03, No. CMS-CSST-2021-B01.
XDL acknowledges the support from the NSFC grant (No. 11803094), 
the Science and Technology Program of Guangzhou, China (No. 202002030360). 
ZQH acknowledges the support from the NSFC grant (No. 12073088).
CGS acknowledges support via the Basic Science Research Program from the National Research Foundation of South Korea (NRF) funded by the Ministry of Education (2018R1A6A1A06024977 and 2020R1I1A1A01073494).


\

\section*{Data availability}

The BigMD simulation used in this paper is availalbe via the CosmoSim database (https://www.cosmosim.org/).
The COLA simulation data used in this paper is available upon request.

\bibliographystyle{aasjournal}

\end{document}